\documentclass[12pt,a4paper]{article}%
\usepackage{amsmath,amssymb,amsthm}
\usepackage{geometry}
\usepackage{amsmath}
\usepackage{amsfonts}
\usepackage{amssymb}
\usepackage{graphicx}%
\providecommand{\U}[1]{\protect\rule{.1in}{.1in}}
\begin{document}

\title{Thermal magnetized D-branes on $\mathbb{R}^{1,p}\times{\mathbb{T}}^{d-p-1}$ in
the generalized Thermo Field Dynamics approach}
\author{R. Nardi\thanks{email: rnardi@cbpf.br}\\\emph{Centro Brasileiro de Pesquisas F\'isicas (CBPF) } \\\emph{R. Dr. Xavier Sigaud 150, 22290-180 Rio de Janeiro - RJ, Brazil}\\M. A. Santos\thanks{email: masantos@cce.ufes.br} \\\emph{Departamento de F{\'i}sica e Q{\'i}mica, } \\\emph{Universidade Federal do Esp{\'i}rito Santo (UFES),} \\\emph{Av. Fernando Ferarri S/N - Goiabeiras, 29060-900
Vit\'{o}ria---ES,Brazil} \\and \\I. V. Vancea\thanks{email: ionvancea@ufrrj.br} \\\emph{Grupo de F{\'{\i}}sica Te\'{o}rica e Matem\'{a}tica F\'{\i}sica,}\\\emph{Departamento de F\'{\i}sica, } \\\emph{Universidade Federal Rural do Rio de Janeiro (UFRRJ),} \\\emph{Cx. Postal 23851, BR 465 Km 7, 23890-000 Serop\'{e}dica - RJ, Brazil }}
\date{9 May 2011}
\maketitle

\begin{abstract}
We construct the D-brane states at finite temperature in thermal equilibrium
in the $\mathbb{R}^{1,p}\times{\mathbb{T}}^{d-p-1}$ spacetime in the presence
of cold (unthermalized) Kalb-Ramond (KR) and $U(1)$ gauge potential
background. To this end, we first generalize the Thermo Field Dynamics (TFD)
to wrapped closed strings. This generalization is consistent with the spatial
translation invariance on the string world-sheet. Next, we determine the
thermal string vacuum and define the entropy operator. From these data we
calculate the entropy of the closed string and the free energy. Finally, we
define the thermal D-brane states in $\mathbb{R}^{1,p}\times{\mathbb{T}%
}^{d-p-1}$ in the presence of cold constant KR field and $U(1)$ gauge
potential as the boundary states of the thermal closed string and compute
their entropy.

\end{abstract}

\newpage


\section{Introduction}

The understanding of D-branes boundary states in $\mathbb{R}^{1,p}%
\times{\mathbb{T}}^{d-p-1}$ spacetime is very important for fundamental
reasons as well as for their applications. The D-branes in $\mathbb{R}%
^{1,p}\times{\mathbb{T}}^{d-p-1}$ are of intrinsic importance for the String
Theory since the background $\mathbb{R}^{1,p}\times{\mathbb{T}}^{d-p-1}$ with
fluxes represent one of the few backgrounds on which the strings can be
defined perturbatively and the D-brane boundary conditions can be solved
exactly. The magnetized D-brane systems in $\mathbb{R}^{1,p}\times{\mathbb{T}%
}^{d-p-1}$ can be used to engineer D-brane configurations that can support
chiral gauge theories in four dimensions \cite{bkls}. The D-branes wrapped
around the cycles of ${\mathbb{T}}^{d-p-1}$ can support gravitons in
$\mathbb{R}^{1,p}$ through the Kaluza-Klein mechanism which is useful to make
contact with the brane-world scenarios. Important results have been obtained
along these lines, from which one could cite: the construction of various
extensions of the standard model with three generations of chiral fermions
from configurations of magnetic D9-branes on orbifolds
\cite{bkl,ml,blt,cau,lg,ad,akp,cln}; the moduli stabilization with fluxes
\cite{bt,akm}; the effect of the instantons and the K\"{a}hler metric on the
chiral multiplets \cite{bfpvlm,mb,vlmf1,vlmf2}. These results establish
relationships between the four dimensional field theory and string theory
phenomenology\footnote{Very important results have been obtained in the low
energy description of magnetized D-branes, too, in which the gravitational
effects can be studied at the expense of neglecting the quantum structure, but
here we will not address the D-branes from this point of view.}. (For more
details on how the extensions of the Standard Model can be obtained from
magnetic D-branes and various examples see \cite{bkls,ia} and the references
therein.) The wrapping of the D-branes on six cycles of the ten dimensional
spacetime represents a key ingredient in these examples. Magnetized maximal
D-brane states on tori were constructed recently in \cite{vlmpp1,drss,vlmpp2}
following the method from \cite{ip1}. (A detailed analysis of the topological
structure of the boundary states on $U(N)$ gauge bundles was done in
\cite{drss} and the T-duality was analysed in \cite{ip2}.)

The existence of the microscopic structure of the magnetized D-branes suggests
that they should have an intrinsic thermodynamics created by the closed string
excitations that condensate on the world-volume. In this paper, we are going
to make the idea of the thermodynamics of the magnetized D-brane more precise
by constructing the thermal D-branes on $\mathbb{R}^{1,p}\times{\mathbb{T}%
}^{d-p-1}$ and calculating their entropy under the hypothesis of the local
thermodynamical equilibrium. The technique to construct thermal D-brane
boundary states from D-branes at zero temperature was developed in
\cite{ivv1,ivv2,ivv3,ivv4,ivv5} and its various aspects were reviewed and
discussed in \cite{ivv6,ivv7,ivv8,ivv9,ivv10}. It is based on the Thermo Field
Dynamics (TFD) approach to field theory at finite temperature which was
previously applied to string theory and string field theory in
\cite{yl1,yl2,yl3,yl4,yl5,yl6,fns1,fn1,fn2,fn3,fn4,fn5,f1} and more recently
in \cite{ng1,ng2,ng3,ng4,ng5}. In this setting, the thermal D-branes are
defined by boundary conditions on the thermal closed string which, on its
turn, is obtained by applying the Bogoliubov operator of all string
oscillators to the closed string at zero temperature\footnote{The thermal
D-branes in the open string channel and their connection with the entangled
string states have been discussed recently in \cite{mbc1,mbc2}.}. The D-brane
equilibrium thermodynamical functions are defined as the expectation values of
the corresponding thermal string operators in the thermal D-brane state. The
statistical properties of the strings in the presence of D-branes have been
discussed in the literature in the path integral formulation in
\cite{vm,amss1,amss2,kh}. However, the TFD approach to the thermodynamics of
D-branes has the advantage that explicit thermal D-brane states can be
constructed in a similar way to the zero temperature theory and statistical
observables can be computed as expectation values of the corresponding
operators in these states. This can be used, for example, to identify the
sectors where the symmetries are broken due to thermal effects (see, for
example, \cite{ivv4}).

In order to obtain the thermal D-brane states in $\mathbb{R}^{1,p}%
\times{\mathbb{T}}^{d-p-1}$, the general method previously developed for
D-branes in $\mathbb{R}^{1,d-1}$ must be modified because of the non-trivial
zero modes that arise from wrapping the closed string on the torus. The zero
mode, or topological sector introduces a temperature dependent factor in the
thermal string vacuum \footnote{The understanding of the winding modes of
strings and branes is of utmost importance in the string cosmology. It has
been shown that the string winding modes lead to a positive Hagedorn
temperature and that they determine the energy of a gas of string and its
pressure \cite{are}. On the other hand, the brane winding modes lead to three
un-compact space-like dimensions by determining the hierarchy in the size of
the extra-dimension without the cosmological horizon problem \cite{bv}.}. We
show that the thermalization of the winding modes is different from that of
the string oscillators and leads to products of the Jacobi theta functions in
the partition function, unlike the products of Bose-Einstein distributions for
the oscillating modes. Consequently, the topological thermal string vacuum and
the topological entropy operator in the ${\mathbb{T}}^{d-p-1}$ subspace are no
longer related to the corresponding objects at zero temperature by
Bogoliubov-like operators. However, we show that the mapping can be put in an
operatorial form by using the formal (Heisenberg) algebra of creation and
annihilation operators for the winding number and the center of mass momentum
along the ${\mathbb{T}}^{d-p-1}$ compact directions.

The intrinsic thermodynamics of the magnetic D-brane can be calculated without
any matter field. However, the background fields in the magnetized brane
models are necessary in order to stabilize the moduli and to cancel out the
tadpole anomalies. Therefore, we consider the general background with the
constant Kalb-Ramond (KR) and the constant $U(1)$ gauge potential. We make the
simplifying assumption that the KR and $U(1)$ fields are cold, that is, the
average values of the corresponding string states are taken to be the same as
at zero temperature. The most general case when the fluxes are thermalized as
a consequence of string thermalization will be studied elsewhere.

The paper is organized as follows. In Section 2 we firstly review the wrapped
magnetized bosonic boundary states in $\mathbb{R}^{1,p}\times{\mathbb{T}%
}^{d-p-1}$ at zero temperature following \cite{vlmpp1}. The topological sector
of these states, i. e. the sector corresponding to the momentum and number
operators along the cycles of the torus, is given at a fixed pair of cycles
$(i,j)$, i. e. the states belong to the subspace $\mathcal{H}_{(i,j)}$ of the
Hilbert space. We generalize this solution to the full Hilbert space by
including all cycles of the torus. Since we are interested in the physical
degrees of freedom only, we choose to work in the light-cone gauge with the
light-cone coordinates from $\mathbb{R}^{1,p}$. In Section 3 we construct the
physical thermal vacuum and the Bogoliubov operators for the bosonic string.
The non-trivial zero modes are a consequence of the wrapping of the closed
string around ${\mathbb{T}}^{d-p-1}$ and of the interaction between the string
and the KR background. We show that the topological terms introduce a
temperature dependent factor in the thermal vacuum but leave the form of the
Bogoliubov operators of the string oscillators unchanged. However, the thermal
parameter is modified with a term proportional to the ${\mathbb{T}}^{d-p-1}$
typical radius. Next, we generalize the TFD method to bosonic string fields on
${\mathbb{T}}^{d-p-1}$ consistently with the spatial translation invariance on
the string world-sheet in $\mathbb{R}^{1,p}\times{\mathbb{T}}^{d-p-1}$. The
mapping between the thermal vacuum and the vacuum at zero temperature is given
in the operatorial form. In the $\mathbb{R}^{1,p}$ sector, the mapping is
reduced to a product of the Bogoliubov operators. In the topological
${\mathbb{T}}^{d-p-1}$ sector we obtain a new operator which we write in the
form of a product of creation and annihilation operators for the winding
number and the center of mass momentum along the compact directions. By using
this construction, we are able to define the entropy operator and calculate
the entropy and the free energy of the closed string in both oscillator and
topological sectors. In Section 4 we construct the thermal magnetized D-brane
states by applying the modified generalized TFD formalism obtained in Section
3 and calculate the D-brane entropy. The last section is devoted to conclusions.

\section{Magnetized D-branes at $T = 0$}

Consider the bosonic closed string in $\mathbb{R}^{1,p}\times{\mathbb{T}%
}^{d-p-1}$ in the presence of the $U(1)$ field $A^{\mu}$ and the constant
Kalb-Ramond field $B_{\mu\nu}=-B_{\nu\mu}$, and let $G_{\mu\nu}=\left\{
\eta_{ab},G_{ij}\right\}  $ be the metric on the target space. Here,
$\eta_{ab}$ is the Minkowski metric on $\mathbb{R}^{1,p}$ and $G_{ij}$ is the
internal metric on the torus ${\mathbb{T}}^{d-p-1}$. Also, we consider for
simplicity the factorization of the gauge field and KR field in to $A^{\mu
}=\left\{  A^{a},A^{i}\right\}  $ and $B_{\mu\nu}=\left\{  B_{ab}%
,B_{ij}\right\}  $, respectively. We are using the following index notation:
$\mu,\nu=\overline{0,d}$ are the target space indices, $a,b=\overline{0,p}$
are $\mathbb{R}^{1,p}$ indices and $i,j=\overline{p+1,d-1}$ are ${\mathbb{T}%
}^{d-p-1}$ indices, respectively. We denote the bosonic string coordinates by
$X^{\mu}(\tau,\sigma)$. Also, since we are interested in the physical degrees
of freedom of the string, we fix the light-cone gauge $X^{+}=\alpha^{\prime
}p_{+}\tau$ in the target space. In the light-cone gauge, $a,b=\overline{2,p}$
are $\mathbb{R}^{p-1}$ indices and $\mu,\nu=\overline{p+1,d-1}$ label
$\mathbb{R}^{p-1}\times{\mathbb{T}}^{d-p-1}$ spacetime objects.

The classical dynamics of the bosonic string fields can be derived from the
following action
\begin{equation}
S_{0}=-\frac{1}{4\pi\alpha^{\prime}}\int d^{2}\sigma\left(  \eta^{\alpha\beta
}G_{\mu\nu}+\epsilon^{\alpha\beta}B_{\mu\nu}\right)  \partial_{\alpha}X^{\mu
}\partial_{\beta}X^{\nu},\label{bos-action}%
\end{equation}
where $\alpha,\beta=0,1$, $\sigma^{\alpha}=(\tau,\sigma)$ and $\epsilon
^{\alpha\beta}$ is the Kronecker symbol in two dimensions with $\epsilon
^{01}=1$. The corresponding equations of motion and closed string boundary
conditions have the form
\begin{align}
\partial^{\alpha}\partial_{\alpha}X^{\mu} &  =0,\label{bosonic-eq-mot}\\
X^{\mu}\left(  \tau,\sigma+\pi\right)   &  =X^{\mu}\left(  \tau,\sigma\right)
+2\pi R^{i}m^{i}\delta^{\mu i},\label{bosonic-bc}%
\end{align}
for all $\mu=\overline{2,d-1}$ and all $i=\overline{p+1,d-1}$. Here, $R^{i}$'s
are the torus radii in the corresponding directions. The solutions of the
equations (\ref{bosonic-eq-mot}) with the boundary conditions
(\ref{bosonic-bc}) have the following Fourier expansion
\begin{align}
X^{a}(\tau,\sigma) &  =x^{a}+2\alpha^{\prime}p^{a}\tau+i\sqrt{\frac
{\alpha^{\prime}}{2}}\sum_{n\neq0}\frac{1}{n}\left[  \hat{a}_{n}%
^{a}e^{-2in(\tau-\sigma)}+\hat{b}_{n}^{a}e^{-2in(\tau+\sigma)}\right]
,\label{sol-1}\\
X^{j}(\tau,\sigma) &  =x^{j}+\sqrt{\alpha^{\prime}}\left[  2\hat{m}^{j}%
\sigma+2G^{jk}\left(  \hat{n}_{k}-B_{kl}\hat{m}^{l}\right)  \tau\right]  \\
&  +i\sqrt{\frac{\alpha^{\prime}}{2}}\sum_{n\neq0}\frac{1}{n}\left[  \hat
{a}_{n}^{j}e^{-2in(\tau-\sigma)}+\hat{b}_{n}^{j}e^{-2in(\tau+\sigma)}\right]
,\label{sol-2}%
\end{align}
where $\hat{n}^{j}$ is the center-of-mass number operator and $\hat{m}^{j}%
\in\mathbb{Z}$ is the winding number operator in the compact direction $j$,
respectively. The eigenvalues of both kinds of operators are entire numbers.
In what follows we are going to normalize the bosonic string operators to the
oscillator operators
\begin{align}
\hat{a}_{m}^{\mu} &  =\sqrt{m}\hat{\alpha}_{m}^{\mu}~~,~~\hat{a}_{-m}^{\mu
}=\sqrt{m}\hat{\alpha}_{m}^{\mu\dagger},\\
\hat{b}_{m}^{\mu} &  =\sqrt{m}\hat{\beta}_{m}^{\mu}~~,~~\hat{b}_{-m}^{\mu
}=\sqrt{m}\hat{\beta}_{m}^{\mu\dagger},
\end{align}
for all $m>0$. The physical Hilbert space is the tensor product of the Hilbert
spaces of all string modes factorized by the level matching condition
\begin{equation}
G_{\mu\nu}\left[  {\hat{n}^{\mu}\hat{m}^{\nu}}+\sum_{n>0}n\left(  \hat{\alpha
}_{n}^{\mu\dagger}\hat{\alpha}_{n}^{\nu}-\hat{\beta}_{n}^{\mu\dagger}%
\hat{\beta}_{n}^{\nu}\right)  \right]  \left\vert \Psi_{phys}\right\rangle
=0,\label{level-matching}%
\end{equation}
where the operators $\hat{n}^{\mu}$ and $\hat{m}^{\mu}$ are zero along the
non-compact directions. The units have been chosen such that the square of the
winding number quantize the string energy in multiples of $\varepsilon
_{s}=l_{s}^{-1}=(\sqrt{\alpha^{\prime}})^{-1}$. The total Hamiltonian is the
sum of $\hat{H}^{\mathbb{R}}$ and $\hat{H}^{{\mathbb{T}}}$. If we choose the
reference frame of the center-of-mass in $\mathbb{R}^{p-1}$, the Hamiltonians
are
\begin{align}
\hat{H}^{\mathbb{R}} &  =\frac{1}{2}\sum_{n\neq0}n:\left(  \hat{\alpha}%
_{n}^{a\dagger}\hat{\alpha}_{n}^{a}+\hat{\beta}_{n}^{a\dagger}\hat{\beta}%
_{n}^{a}\right)  :,\label{Hamiltonian-R}\\
\hat{H}^{{\mathbb{T}}} &  =\frac{1}{2\sqrt{\alpha^{\prime}}}\left[  G_{ij}%
\hat{m}^{i}\hat{m}^{j}+\left(  \hat{n}_{i}-B_{ik}\hat{m}^{k}\right)
G^{ij}\left(  \hat{n}_{j}-B_{jl}\hat{m}^{l}\right)  \right]  +\frac{1}%
{2}G_{ij}\sum_{n\neq0}n:\left(  \hat{\alpha}_{n}^{i\dagger}\hat{\alpha}%
_{n}^{j}+\hat{\beta}_{n}^{i\dagger}\hat{\beta}_{n}^{j}\right)
:.\label{Hamiltonian-T}%
\end{align}
As observed in \cite{vlmpp2}, in order to obtain the correct number of degrees
of freedom of the supersymmetric boundary state in the compact directions, one
must take $\hat{H}^{{\mathbb{T}}}/2$ in the total Hamiltonian.

The bosonic magnetized D-brane boundary state $\left\vert B\right\rangle $ is
defined by the boundary conditions in the closed string Hilbert space. The
zero mode boundary conditions are given by the following equations%
\begin{equation}
\hat{p}^{a}\left\vert B\right\rangle =0~,~~\left(  \hat{n}_{i}-2\pi
\alpha^{\prime}qF_{ij}\hat{m}^{j}\right)  \left\vert B\right\rangle =0,
\label{bc-R-T-momentum}%
\end{equation}
where $F_{\mu\nu}=\partial_{\mu}A_{\nu}-\partial_{\nu}A_{\mu}$. The string
fields couple with the gauge invariant combination of the $U(1)$ field and the
KR field
\begin{equation}
\mathcal{B}_{\mu\nu}=\left(  B-2\pi\alpha^{\prime}qF\right)  _{\mu\nu}.
\label{invariant-B}%
\end{equation}
The invariance of the gauge field under translations along the cocycles of the
torus imply that the the components of $F$ are integers \cite{gth,vlmpp2}. The
coupling determines the oscillator boundary conditions in terms of creation
and annihilation operators%
\begin{align}
\left[  \left(  \mathbf{1}+\mathcal{B}\right)  _{ab}\hat{\alpha}_{n}%
^{b}+\left(  \mathbf{1}+\mathcal{B}\right)  _{ab}^{T}\hat{\beta}_{n}^{b\dag
}\right]  \left\vert B\right\rangle  &  =0,\label{bc-R-boson-1}\\
\left[  \left(  \mathbf{1}+\mathcal{B}\right)  _{ab}\hat{\alpha}_{n}^{b\dag
}+\left(  \mathbf{1}+\mathcal{B}\right)  _{ab}^{T}\hat{\beta}_{n}^{b}\right]
\left\vert B\right\rangle  &  =0,\label{bc-R-boson-2}\\
\left(  \mathcal{E}_{ij}\hat{\beta}_{n}^{j}+\mathcal{E}_{ij}^{T}\hat{\alpha
}_{n}^{j\dag}\right)  \left\vert B\right\rangle  &  =0,\label{bc-T-boson-1}\\
\left(  \mathcal{E}_{ij}\hat{\beta}_{n}^{j\dag}+\mathcal{E}_{ij}^{T}%
\hat{\alpha}_{n}^{j}\right)  \left\vert B\right\rangle  &  =0,
\label{bc-T-boson-2}%
\end{align}
where $n>0$ is the mode index and $\mathcal{E}_{ij}=\left(  G-\mathcal{B}%
\right)  _{ij}$. The solution to the equations (\ref{bc-R-boson-1}%
)-(\ref{bc-R-boson-2}) can be factorized as:
\begin{equation}
\left\vert B\right\rangle =\left\vert B\right\rangle _{osc}^{\mathbb{R}%
}\otimes\left\vert B\right\rangle _{osc}^{{\mathbb{T}}}\otimes\left\vert
B\right\rangle _{top}^{{\mathbb{T}}}. \label{factorized-solution}%
\end{equation}
The states $\left\vert B\right\rangle _{osc}^{\mathbb{R}}$ and $\left\vert
B\right\rangle _{osc}^{{\mathbb{T}}}$ are generated by oscillators only, and
their form is already known \cite{vl1,vl2}
\begin{align}
\left\vert B\right\rangle _{osc}^{\mathbb{R}}  &  =N_{p-1}\left(
\mathcal{B}\right)  \left(  {\displaystyle\prod\limits_{n=1}^{\infty}}%
e^{-\hat{\alpha}_{n}^{a\dag}M_{ab}\hat{\beta}_{n}^{b\dag}}\right)  \left\vert
0\right\rangle _{osc},\label{bosonic-B-R-osc}\\
\left\vert B\right\rangle _{osc}^{{\mathbb{T}}}  &  =N_{d-p-1}\left(
\mathcal{B}\right)  \left(  {\displaystyle\prod\limits_{n=1}^{\infty}}%
e^{-\hat{\alpha}_{n}^{i\dag}G_{ik}\left(  \mathcal{E}^{-1}\right)
^{kl}\left(  \mathcal{E}^{T}\right)  _{lj}\hat{\beta}_{n}^{j\dag}}\right)
\left\vert 0\right\rangle _{osc}, \label{bosonic-B-T-osc}%
\end{align}
where the constants $N_{p-1}\left(  \mathcal{B}\right)  $ and $N_{d-p-1}%
\left(  \mathcal{B}\right)  $ depend on the background fields and brane
tension and can be computed from the cilinder diagram interpreted in the open
and closed string sectors. The topological boundary state $\left\vert
B\right\rangle _{top}^{{\mathbb{T}}}$ is a solution of the second equation
from (\ref{bc-R-T-momentum}). As can be easily checked, it has the follwing
form
\begin{equation}
\left\vert B\right\rangle _{top}^{{\mathbb{T}}}=\prod_{i=p+1}^{d-1}\sum
_{n_{i},m_{i}}\delta_{n_{i}-2\pi\alpha^{\prime}qF_{i}^{k_{i}}m_{k_{i}}%
}\left\vert n_{i}\right\rangle \left\vert m_{i}\right\rangle .
\label{bosonic-B-T-top}%
\end{equation}
In what follows, we are going to use the following normalization of momentum
and wrapping number eigenstates
\begin{equation}
\left\langle p^{a}|p^{\prime b}\right\rangle =2\pi\delta^{ab}\delta\left(
p-p^{\prime}\right)  ,~\left\langle n_{i}|n^{\prime}{}_{j}\right\rangle
=\left(  2\pi\sqrt{\alpha^{\prime}}\right)  ^{\frac{1}{2}}\delta
_{n_{i},n^{\prime}{}_{j}},~\left\langle m^{i}|m^{\prime j}\right\rangle
=\left(  2\pi\sqrt{\alpha^{\prime}}\right)  ^{\frac{1}{2}}\delta
_{m^{i},m^{\prime j}}. \label{normalization-momentum}%
\end{equation}
Note that the state given in (\ref{bosonic-B-T-top}) generalizes the one given
in \cite{vlmpp1} that satisfies the equation (\ref{bc-R-T-momentum}) only in
two directions of the torus. The maximal flat D-brane boundary state is
obtained for $p=d-1$.

\section{Thermodynamics of closed string in $\mathbb{R}^{p-1}\times
{\mathbb{T}}^{d-p-1}$}

The boundary states from the equation (\ref{factorized-solution}) describe
condensates of the string modes in $\mathbb{R}^{p-1}\times{\mathbb{T}}%
^{d-p-1}$. Our main goal is to calculate the thermal magnetized D-branes and
study their thermodynamical properties. To this end, we need to generalize the
method from \cite{ivv1,ivv2,ivv4} which has been used to study the thermal
D-branes in flat spacetime to $\mathbb{R}^{p-1}\times{\mathbb{T}}^{d-p-1}$.
This method already generalizes the TFD to closed string boundary states and
it is the only one that provides the explicit construction of the thermal
vacuum states and thermal boundary states \footnote{Note that in the path
integral treatment of the strings at finite temperature the thermal vacuum and
the D-brane states are not explicitly constructed.}.

In the TFD approach, one studies the thermodynamical properties of the
D-branes in the flat spacetime by allowing the system to interact with a
thermal reservoir. This interaction, called \emph{thermalization}, is
described in terms of the Bogoliubov operator that acts on the total string
Hilbert space which is the tensor product of the closed string Hilbert space
and the Hilbert space of an identical copy of the original string denoted by a
tilde. The tilde string describes the reservoir degrees of freedom that
interact, one by one, with the string degrees of freedom \cite{ubook}. The
Bogoliubov operator maps the total Hilbert space to the thermal Hilbert space
which has a Fock space structure, that is, the thermal states can be obtained
by acting with thermal creation and annihilation operators on the physical
thermal vacuum $|0\left(  \beta_{T}\right)  \rangle\rangle$ \cite{ubook}.
However, in the case of the thermal magnetized D-brane in $\mathbb{R}%
^{p-1}\times{\mathbb{T}}^{d-p-1}$, there is an extra topological sector of the
string Hilbert space as reviewed in the previous section. There is no TFD
prescription for the thermalization of the topological modes and their
inclusion in to the thermal vacuum state. Therefore, our main goal in this
section is to generalize the TFD method to the string in $\mathbb{R}%
^{p-1}\times{\mathbb{T}}^{d-p-1}$ and to determine the thermal operator that
performs the thermalization of the topological sector. We will calculate the
physical thermal vacuum of the closed string in the presence of the constant
$U(1)$ field and the Kalb-Ramond field from the principles of the TFD.

\subsection{Thermal vacuum of closed string}

In the TFD, the thermal vacuum of a general system at the thermodynamical
equilibrium is defined by postulating that for any observable $\widehat{A}$
the statistical average calculated with the canonical ensemble is the vacuum
expectation value of $\widehat{A}$ in the thermal vacuum%
\begin{equation}
\left\langle \widehat{A}\right\rangle =Z^{-1}\left(  \beta_{T}\right)
Tr\left[  \widehat{A}e^{-\beta_{T}\widehat{H}}\right]  =\left\langle
\left\langle 0\left(  \beta_{T}\right)  \left\vert \widehat{A}\right\vert
0\left(  \beta_{T}\right)  \right\rangle \right\rangle , \label{thermal-vac}%
\end{equation}
where $\beta_{T}=1/k_{B}T\,$\footnote{If the Lorentz invariant product
$\beta_{T}P^{0}$ is used to define the canonical ensemble, then the trace from
(\ref{thermal-vac}) should be replaced by $\int dp^{+}Tr[\hat{A}\exp
({-\frac{\beta_{T}}{p^{+}}\widehat{H}})]$ (see e.g. \cite{ao,agn}).}.

The thermal string vacuum in $\mathbb{R}^{p-1}\times{\mathbb{T}}^{d-p-1}$ can
be obtained by generalizing this general TFD as follows. Since in the general
definition (\ref{thermal-vac}) the trace is taken over all physical states of
the system under consideration, in the case of string theory the relation
(\ref{thermal-vac}) must be modified in order to produce a thermal vacuum from
physical states only. The modification consists in imposing the constraint
that the trace be taken over the states that are invariant under the
translation of the worldsheet along the $\sigma$ direction (the level matching
condition). This condition is necessary and sufficient in the case of the
bosonic string and the GS superstring \cite{ivv1,ivv4,ivv5,ng3,ng5}. It
follows that, in order to construct the thermal vacuum from physical states,
the relation (\ref{thermal-vac}) should be generalized to the following
relations%
\begin{align}
\left\langle \hat{A}\right\rangle  &  =Z^{-1}\left(  \beta_{T}\right)
\sqrt{\alpha^{\prime}}\int_{0}^{1}d\lambda Tr\left[  e^{-\beta_{T}\hat
{H}\left(  p^{a},N,M,N_{osc}\right)  +2\pi i\lambda\hat{P}}\hat{A}\right]
\nonumber\\
&  =\sqrt{\alpha^{\prime}}\int d^{p-2}p\int_{0}^{1}d\lambda\left\langle
\left\langle 0\left(  \beta_{T},\lambda,p^{a}\right)  \left\vert \delta\left(
\hat{P}=0\right)  \hat{A}\right\vert 0\left(  \beta_{T},\lambda,p^{a}\right)
\right\rangle \right\rangle . \label{thermal-vac-string}%
\end{align}
The action of the string momentum $\hat{P}=\hat{L}_{0}^{l}-\hat{L}_{0}^{r}$ on
the states is defined by the relation (\ref{level-matching}). The trace over
$\hat{P}$ involves the integral over the real Lagrange multiplier $\lambda$
and the sum over the indices that label vectors from the physical subspace of
the total Hilbert space \cite{ivv4,ng1}. The same integral should be taken in
the r.h.s. of the relation (\ref{thermal-vac-string}) where the thermal string
vacuum dependes on the variable $\lambda$ \footnote{It has been shown in
\cite{ng1} that the ambiguity in choosing an unitary Bogoliubov transformation
for strings in TFD can be lifted if one requires that the transformation
minimizes the free energy of the form $F(\theta)=\int_{0}^{1}d\lambda
f(\theta,\lambda)$. As a consequence, the Bogoliubov transformation depends on
$\lambda$ which implies that the thermal vacuum depends on $\lambda$, too.}.
The string Hamiltonian $\hat{H}\left(  p^{a},N,M,N_{osc}\right)  $ is given by
the sum of the two operators from (\ref{Hamiltonian-R}) and
(\ref{Hamiltonian-T}). The string partition function is the trace of the
identity operator
\begin{equation}
Z\left(  \beta_{T}\right)  =\sqrt{\alpha^{\prime}}\int d^{p-2}p\int_{0}%
^{1}d\lambda Tr\left[  e^{-\beta_{T}\hat{H}\left(  p^{a},N,M,N_{osc}\right)
+2\pi i\lambda\hat{P}}\right]  . \label{part-function}%
\end{equation}
The thermal string vacuum $\left\vert \left.  0\left(  \beta_{T},\lambda
,p^{a}\right)  \right\rangle \right\rangle $ belongs to the total Hilbert
space%
\begin{equation}
\mathcal{H}_{phys}^{tot}=\mathcal{H}_{phys}\otimes\widetilde{\mathcal{H}%
}_{phys}. \label{total-physical-hilbert-space}%
\end{equation}
After the projection onto the physical Hilbert subspace, one is left with the
physical thermal vacuum $\left\vert \left.  0\left(  \beta_{T},p^{a}\right)
\right\rangle \right\rangle $. The projections of $\mathcal{H}$ onto
$\mathcal{H}_{phys}$ and of $\widetilde{\mathcal{H}}$ onto $\widetilde
{\mathcal{H}}_{phys}$ are given by the level matching condition for the string
and the tilde string, respectively. In order to simplify the notation, we have
introduced the following multi-indices: $N$ and $M$ represent all compact
linear momenta and all winding numbers, respectively, while $N_{osc}$ is the
number of all left- and right-moving string oscillators along all directions
from $\mathbb{R}^{p-1}$ and ${\mathbb{T}}^{d-p-1}$
\begin{equation}
\left\vert N\right\rangle =\bigotimes\limits_{i=1}^{d-p-1}\left\vert
n_{i}\right\rangle ,\left\vert M\right\rangle =\bigotimes\limits_{i=1}%
^{d-p-1}\left\vert m^{i}\right\rangle ,\left\vert N_{osc}\right\rangle
=\bigotimes\limits_{n_{1},n_{2},\ldots=1}^{\infty}\left\vert n_{1}%
,n_{2},\ldots\right\rangle , \label{string-products}%
\end{equation}
where%
\begin{equation}
\widehat{n}_{i}\left\vert n_{i}\right\rangle =n_{i}\left\vert n_{i}%
\right\rangle ,~~\widehat{m}^{i}\left\vert m^{i}\right\rangle =m^{i}\left\vert
m^{i}\right\rangle . \label{eigenstates-N-M}%
\end{equation}
The string states are tensor products of the form%
\begin{equation}
\left\vert N,M,N_{osc}\right\rangle =\left\vert N\right\rangle \left\vert
M\right\rangle \left\vert N_{osc}\right\rangle . \label{string-states}%
\end{equation}
The state $|0\left(  \beta_{T},p^{a}\right)  \rangle\rangle$ can be obtained
from the second equality of (\ref{thermal-vac}). In order to compute the
integral over $\lambda$ from the trace, one has to choose an explicit form of
the metric. In what follows, we choose ${\mathbb{T}}^{d-p-1}$ to be flat with
the metric $G_{\mu\nu}=\left(  \delta_{ab},\frac{R^{2}}{\alpha^{\prime}}%
\delta_{ij}\right)  $, where $R^{i}=R$ is the typical compactification radius.
Without loss of generality, we can work in $\mathbb{R}^{p-1}$ in the reference
frame of the string center of mass in which $|0\left(  \beta_{T},p^{a}\right)
\rangle\rangle|_{c.m.}=$ $|0\left(  \beta_{T}\right)  \rangle\rangle$. After a
simple algebra, the relation (\ref{thermal-vac}) can be cast in to the
following form
\begin{align}
&  \left\langle \left\langle 0\left(  \beta_{T}\right)  \left\vert
\delta\left(  \widehat{P}=0\right)  \widehat{A}\right\vert 0\left(  \beta
_{T}\right)  \right\rangle \right\rangle =Z^{-1}\left(  \beta_{T}\right)
\sqrt{\alpha^{\prime}}\nonumber\\
&  \times\sum_{N,M,N_{osc}}e^{-\beta_{T}\left[  E(N,M,N_{osc})\right]  }%
\int_{0}^{1}d\lambda e^{2\pi i\lambda P}A_{N,M,N_{osc};N,M,N_{osc}},
\label{thermal-vac-1}%
\end{align}
The notation $A_{N,M,N_{osc};N,M,N_{osc}}$stands for the matrix elements of
$\hat{A}$ in the corresponding string states. The eigenvalues of the
Hamiltonian from (\ref{thermal-vac-1}) have the following form%
\begin{equation}
E\left(  N,M,N_{osc}\right)  =\frac{R^{2}}{2(\alpha^{\prime})^{\frac{3}{2}}%
}\left[  \left(  m\right)  ^{2}+\left(  n-Bm\right)  ^{2}\right]  +\frac{1}%
{2}\sum_{n>0}\left(  E_{l,n}^{\mathbb{R}}+E_{r,n}^{\mathbb{R}}\right)
+\frac{R^{2}}{2\alpha^{\prime}}\sum_{n>0}\left(  E_{l,n}^{{\mathbb{T}}%
}+E_{r,n}^{{\mathbb{T}}}\right)  , \label{E-number-total}%
\end{equation}
where the energy of the left- and right-moving oscillators is $E_{\ast
}=nN_{\ast,n}$ for $\ast=l,r$. The energy of zero modes is given by the first
term in the above relation. The relation (\ref{thermal-vac-1}) defines the
physical thermal vacuum and it must hold for any observable $\hat{A}$.

The thermal vacuum can be obtained from the equation (\ref{thermal-vac-1}) as
follows. To begin with, we note that the integral over $\lambda$ represents
the integral form of the multivariable $\delta$-function and can be written in
terms of the orthonormal functions%
\begin{equation}
\Psi_{E}(\lambda)=\frac{1}{\sqrt{2\pi}}e^{-2\pi iE\lambda},~~\int_{0}%
^{1}d\lambda\Psi_{E^{\prime}}^{\ast}(\lambda)\Psi_{E}(\lambda)=\frac{1}{2\pi
}\delta_{E^{\prime},E}. \label{orthonormal-functions}%
\end{equation}
When these relations are applied to the left- and right-moving modes, they
lead to the level matching condition which should appear in the matrix
elements of $\hat{A}$ from the r. h. s. of (\ref{thermal-vac-1}) after the
integration over $\lambda$. Next, we decompose the thermal vacuum in the Fock
basis tensored with the compact momentum and widing number basis of the total
physical space \cite{ubook},
\begin{equation}
\left\vert \left.  0\left(  \beta_{T}\right)  \right\rangle \right\rangle
=\sum\limits_{N,M,N_{osc}}\left[  f\left(  \beta_{T}\right)  \right]
_{N,M,N_{osc}}\left\vert N,M,N_{osc}\right\rangle \widetilde{\left\vert
N,M,N_{osc}\right\rangle }, \label{f-coeff}%
\end{equation}
where $\left[  f\left(  \beta_{T}\right)  \right]  _{N,M,N_{osc}}$ are complex
coefficients. By substituting (\ref{f-coeff}) into (\ref{thermal-vac-1}), we
can show that the thermal vacuum has the following form%
\begin{equation}
\left\vert \left.  0\left(  \beta_{T}\right)  \right\rangle \right\rangle
=Z^{-\frac{1}{2}}\left(  \beta_{T}\right)  \sum\limits_{N,M,N_{osc}}%
\exp\left[  -\frac{\beta_{T}}{2}E\left(  N,M,N_{osc}\right)  \right]
\left\vert N,M,N_{osc}\right\rangle \widetilde{\left\vert N,M,N_{osc}%
\right\rangle }. \label{thermal-vac-3}%
\end{equation}
The partition function can be obtained from (\ref{part-function}) and the
normalization of the thermal vacuum to unity. In an arbitrary inertial frame,
$Z\left(  \beta_{T}\right)  $ can be factorized as%

\begin{equation}
Z\left(  \beta_{T}\right)  =Z_{0}\left(  \beta_{T}\right)  Z_{top}\left(
\beta_{T}\right)  Z_{osc}\left(  \beta_{T}\right)  , \label{factor-part-funct}%
\end{equation}
where
\begin{align}
Z_{0}\left(  \beta_{T}\right)   &  =\int d^{p-2}k\left\langle k\left\vert
\exp\left(  -\frac{\beta_{T}}{2}\alpha^{\prime}\hat{p}^{2}\right)  \right\vert
k\right\rangle ,\label{part-funct-0}\\
Z_{top}\left(  \beta_{T}\right)   &  =\sum_{N,M}\left\langle N,M\left\vert
\exp\left\{  -\beta_{T}\frac{R^{2}}{2(\alpha^{\prime})^{\frac{3}{2}}}\left[
\left(  \hat{m}\right)  ^{2}+\left(  \hat{n}-B\hat{m}\right)  ^{2}\right]
\right\}  \right\vert N,M\right\rangle ,\label{part-funct-top}\\
Z_{osc}\left(  \beta_{T}\right)   &  =\sum_{N_{osc}}\left\langle
N_{osc}\left\vert \exp\left[  -\frac{\beta_{T}}{2}\sum_{n>0}n\left(  \hat
{N}_{l,n}+\hat{N}_{r,n}\right)  \right]  \right\vert N_{osc}\right\rangle
\nonumber\\
&  +\int_{0}^{1}d\lambda\sum_{N}\sum_{N_{osc}}\left\langle N_{osc}\left\vert
\exp\left\{  \pi i\lambda\left[  \sum_{n>0}n\left(  \hat{N}_{l,n}-\hat
{N}_{r,n}\right)  +\frac{R^{2}}{(\alpha^{\prime})^{\frac{3}{2}}}\hat
{W}\right]  \right\}  \right\vert N_{osc}\right\rangle ,
\label{part-funct-osc}%
\end{align}
where $N_{osc}=(N_{osc}^{\mathbb{R}},N_{osc}^{\mathbb{T}})$, $\hat{N}%
_{l,n}=\hat{N}_{l,n}^{\mathbb{R}}+\frac{R^{2}}{\alpha^{\prime}}\hat{N}%
_{l,n}^{{\mathbb{T}}}$ and $\hat{N}_{r,n}=\hat{N}_{r,n}^{\mathbb{R}}%
+\frac{R^{2}}{\alpha^{\prime}}\hat{N}_{r,n}^{{\mathbb{T}}}$ and where $\hat
{W}=\sum_{i,j}\delta_{ij}\hat{n}^{i}\hat{m}^{j}$. The factor $Z_{0}\left(
\beta_{T}\right)  $ is a gaussian integral over the momenta of the center of
mass in the $\mathbb{R}^{p-1}$ subspace. In the center of mass reference frame
$Z_{0}\left(  \beta_{T}\right)  $ does not contribute to the partition
function. The integral over $\lambda$ from $Z_{osc}\left(  \beta_{T}\right)  $
can be easily computed by using the orthonormal functions from
(\ref{orthonormal-functions}). The sums over $N=(N_{l,n}^{\mathbb{R}}%
,N_{r,n}^{\mathbb{R}},N_{l,n}^{{\mathbb{T}}}N_{r,n}^{{\mathbb{T}}})$ reduce to
sums over the left-moving oscillator numbers as a consequence of the level
matching condition%
\begin{align}
\mathbb{R\colon\qquad}N_{osc}^{l,a}  &  =N_{osc}^{r,,a}, \label{level-mat-N-R}%
\\
{\mathbb{T\colon\qquad}}N_{osc}^{l,j}  &  =N_{osc}^{r,j}+\frac{12R^{2}%
W}{(d-p-1)(\alpha^{\prime})^{\frac{3}{2}}}, \label{level-mat-N-T}%
\end{align}
for $d>p+1$. For the maximal flat subspace $d=p+1$ it follows that $W=0$ since
there are no compact directions at all. Then it is easy to see that
$Z_{osc}\left(  \beta_{T}\right)  $ has the following form%
\begin{equation}
Z_{osc}\left(  \beta_{T}\right)  =%
{\displaystyle\prod\limits_{n=1}^{\infty}}
\left(  1-e^{-\beta_{T}n}\right)  ^{1-p}(1-e^{-\frac{\beta_{T}R^{2}}%
{2\alpha^{\prime}}n})^{p+1-d}. \label{part-funct-osc-1}%
\end{equation}
The topological partition function $Z_{top}\left(  \beta_{T}\right)  $ can be
calculated, too, and one can show that it is given by the relation%
\begin{align}
Z_{top}\left(  \beta_{T}\right)   &  \sim\sum_{M}\prod_{j=p+1}^{d-1}%
\vartheta\left(  -\frac{i\beta_{T}R^{2}}{4\pi(\alpha^{\prime})^{\frac{3}{2}}%
}C_{k}^{j}m^{k};\frac{i\beta_{T}R^{2}}{2\pi(\alpha^{\prime})^{\frac{3}{2}}%
}\left[  \frac{2(d-p)+22}{d-p-1}\right]  \right) \nonumber\\
&  \times\exp\left\{  -\frac{\beta_{T}R^{2}}{2\pi(\alpha^{\prime})^{\frac
{3}{2}}}\left[  m_{j}^{2}+\left(  B_{jk}m^{k}\right)  ^{2}\right]  \right\}  ,
\label{top-partition-function-1}%
\end{align}
where $C_{j}^{i}=\delta_{j}^{i}+B_{j}^{i}$ and $\vartheta(z;\tau)$ is the
Jacobi's theta function. Note that $\operatorname{Im}\tau>0$ in either $d=10$
or $d=26$ for $d>p+1$ as required by the definition of the $\vartheta
$-function. By collecting the above results and by taking into account the
normalization relations for the momenta (\ref{normalization-momentum}), one
can see that the thermal vacuum is given by the relation%
\begin{align}
|\left.  0\left(  \beta_{T}\right)  \right\rangle \rangle &  =(2\pi
)^{\frac{3-d}{2}}(\alpha^{\prime})^{\frac{3-2p}{4}}R^{\frac{p+1-d}{2}}%
\beta_{T}^{\frac{p-2}{2}}%
{\displaystyle\prod\limits_{n=1}^{\infty}}
\left[  \left(  1-e^{-\beta_{T}n}\right)  ^{1-p}(1-e^{-\frac{\beta_{T}R^{2}%
}{2\alpha^{\prime}}n})^{p+1-d}\right] \nonumber\\
&  \times\left\{  \sum_{M}\prod_{j=p+1}^{d-1}\vartheta\left(  -\frac
{i\beta_{T}R^{2}}{4\pi(\alpha^{\prime})^{\frac{3}{2}}}C_{k}^{j}m^{k}%
;\frac{i\beta_{T}R^{2}}{2\pi(\alpha^{\prime})^{\frac{3}{2}}}\left[
\frac{2(d-p)+22}{d-p-1}\right]  \right)  \right. \nonumber\\
&  \left.  \times\exp\left\{  -\frac{\beta_{T}R^{2}}{2\pi(\alpha^{\prime
})^{\frac{3}{2}}}\left[  m_{j}^{2}+\left(  B_{jk}m^{k}\right)  ^{2}\right]
\right\}  \right\}  ^{-\frac{1}{2}}\nonumber\\
&  \times\sum\limits_{N,M,N_{osc}}\exp\left[  -\frac{\beta_{T}}{2}E\left(
N,M,N_{osc}\right)  \right]  \left\vert N,M,N_{osc}\right\rangle
\widetilde{\left\vert N,M,N_{osc}\right\rangle }. \label{thermal-vac-final}%
\end{align}
Some comments are in order now. One way to interpret the relation
(\ref{part-funct-osc-1}) is by recalling that the energy of the $n$-th mode is
$\varepsilon_{n}=n$ in flat spacetime. Then one can see that the oscillators
from ${\mathbb{T}}^{d-p-1}$ have the energy
\begin{equation}
\varepsilon_{n}^{\prime}=\frac{R^{2}n}{2\alpha^{\prime}},\quad\forall n>0.
\label{energy-osc-T}%
\end{equation}
Another interpretation of the relation (\ref{part-funct-osc-1}) is that the
string modes on the torus behave as they would be at an effective temperature
$T^{\prime}=2\alpha^{\prime}T/R^{2}$. Of course that this is not true, since
the full string should be at the thermodynamical equilibrium. However, this
remark is important since the first major problem of determining the
thermalization of the the string modes in $\mathbb{R}^{p-1}\times{\mathbb{T}%
}^{d-p-1}$, namely the thermalization of the string oscillators, is solved by
the relation (\ref{thermal-vac-final}). Indeed, from it one can see that the
mapping of the string modes to the finite temperature is realized by
Bogoliubov operators. Along the directions of the $\mathbb{R}^{p-1}$, the
Bogoliubov operators are the same as in the flat spacetime
\begin{align}
\hat{G}(\beta_{T})  &  =\sum_{n=1}^{\infty}\sum_{\mu=2}^{d-1}\hat{G}_{n}^{\mu
}(\beta_{T}),\qquad\qquad\hat{G}_{n}^{\mu}(\beta_{T})=\hat{G}_{l,n}^{\mu
}(\beta_{T})+\hat{G}_{r,n}^{\mu}(\beta_{T}),\label{left-right-Bogoliubov-op}\\
\hat{G}_{l,n}^{\mu}(\beta_{T})  &  =-i\theta_{n}(\beta_{T})\left(  \hat
{\alpha}_{n}^{\mu\dagger}\widehat{\tilde{\alpha}}_{n}^{\mu\dagger}-\hat
{\alpha}_{n}^{\mu}\widehat{\tilde{\alpha}}_{n}^{\mu}\right)  ,\quad\hat
{G}_{r,n}^{\mu}(\beta_{T})=-i\theta_{n}(\beta_{T})\left(  \hat{\beta}_{n}%
^{\mu\dagger}\widehat{\tilde{\beta}}_{n}^{\mu\dagger}-\hat{\beta}_{n}^{\mu
}\widehat{\tilde{\beta}}_{n}^{\mu}\right)  . \label{bosonic-Bogoliubov-op}%
\end{align}
The function $\theta_{n}(\beta_{T})$ depends on the temperature and on the
oscillator frequency. Thus, it is the same in $\mathbb{R}^{p-1}$ and
${\mathbb{T}}^{d-p-1}$ and is given by the following equation%
\begin{equation}
\cosh\theta_{n}(\beta_{T})=\left(  1-e^{-\beta_{T}n}\right)  ^{-1/2}.
\label{bosonic-theta}%
\end{equation}
However, for the ${\mathbb{T}}^{d-p-1}$ submanifold, the argument of
$\theta_{n}(\beta_{T})$ should be replaced by $\theta_{n}(\beta_{T}%
R^{2}/2\alpha^{\prime})$. The form of the function $\theta_{n}(\beta_{T})$
given by the relation (\ref{bosonic-theta}) remains unchanged under the
multiplication of the temperature variable by $R^{2}/2\alpha^{\prime}$. It
follows that, in order to thermalize the string modes in $\mathbb{R}%
^{p-1}\times{\mathbb{T}}^{d-p-1}$, one should apply the same method as in the
flat spacetime with the only difference of the rescaled temperature in the
compact subspace.

The second important problem of the thermal string in $\mathbb{R}^{p-1}%
\times{\mathbb{T}}^{d-p-1}$ is the thermalization of the topological states,
which is a process that is not defined by the TFD method. Indeed, in the TFD
approach, the thermalization relies on the Fock space structure of the
oscillator Hilbert space while the topological Hilbert subspace does not have
this structure. From (\ref{thermal-vac-final}), we can see that the
topological sector is not mapped to the finite temperature by a Bogoliubov
operator. In fact, one can factorize the thermal vacuum as%
\begin{equation}
\left\vert \left.  0\left(  \beta_{T}\right)  \right\rangle \right\rangle
\sim\beta_{T}^{\frac{p-2}{2}}\left\vert \left.  0\left(  \beta_{T}\right)
\right\rangle \right\rangle _{top}\sum\limits_{N_{osc}}e^{-\frac{\beta_{T}}%
{2}E(N_{osc})}\left\vert N_{osc}\right\rangle \widetilde{\left\vert
N_{osc}\right\rangle }. \label{thermal-vacuum-factorized}%
\end{equation}
In order to find the map between $\left\vert \left.  0\right\rangle
\right\rangle _{top}$ and $\left\vert \left.  0\left(  \beta_{T}\right)
\right\rangle \right\rangle _{top}$ we introduce the following algebra
$\left\{  \hat{n}_{i},\hat{n}_{i}^{+},\hat{n}_{i}^{-}\right\}  $%
\begin{equation}
\left[  \hat{n}_{i}^{+},\hat{n}_{j}^{-}\right]  =0,\quad\left[  \hat{n}%
_{i},\hat{n}_{j}^{+}\right]  =\hat{n}_{i}^{+}\delta_{ij},\quad\left[  \hat
{n}_{i},\hat{n}_{j}^{-}\right]  =-\hat{n}_{i}^{-}\delta_{ij},
\label{algebra-n-m}%
\end{equation}
with
\begin{equation}
\left(  \hat{n}_{i}^{+}\right)  ^{\dag}=\hat{n}_{i}^{-},\quad\left(  \hat
{n}_{i}^{-}\right)  ^{\dag}=\hat{n}_{i}^{+}. \label{algebra-n-m-property}%
\end{equation}
The action of the above operators on the states $\left\{  \left\vert
n_{i}\right\rangle \right\}  $ is given by the following relations%
\begin{align}
\hat{n}_{i}\left\vert n_{i}\right\rangle  &  =n_{i}\left\vert n_{i}%
\right\rangle ,\label{algebra-n-m-1}\\
\hat{n}_{i}^{+}\left\vert n_{i}\right\rangle  &  =\left\vert n_{i}%
+1\right\rangle ,\label{algebra-n-m-2}\\
\hat{n}_{i}^{-}\left\vert n_{i}\right\rangle  &  =\left\vert n_{i}%
-1\right\rangle . \label{algebra-n-m-3}%
\end{align}
The operators $\hat{n}_{i}^{+}$ and $\hat{n}_{i}^{-}$ create and annihilate
quanta of the momentum of the center of mass along the compact direction $i$
with arbitrary eigenvalues $n_{i}\in%
\mathbb{Z}
$. An identical algebra $\left\{  \hat{m}_{i},\hat{m}_{i}^{+},\hat{m}_{i}%
^{-}\right\}  $ can be introduce to describe the increase and the decrease of
the winding number, with the same unbounded eigenvalues \footnote{A concrete
representation of the algebra $\left\{  \hat{n}_{i},\hat{n}_{i}^{+},\hat
{n}_{i}^{-}\right\}  $ can be given in terms of the wave functions
$\left\langle \sigma|n_{j}\right\rangle =\varphi_{n_{j}}(\sigma)=\exp
(in_{j}\sigma)$ with $\hat{n}_{j}=-i\partial_{\sigma},\hat{n}_{j}^{+}%
=\exp(i\sigma)$ and $\hat{n}_{j}^{-}=\exp(-i\sigma).$}. Then the topological
vacuum can be written as%
\begin{equation}
\left\vert \left.  0\left(  \beta_{T}\right)  \right\rangle \right\rangle
_{top}=\hat{\Omega}\left(  \beta_{T}\right)  \left\vert \left.  0\right\rangle
\right\rangle _{top}, \label{topological-vacuum-map}%
\end{equation}
where
\begin{equation}
\hat{\Omega}\left(  \beta_{T}\right)  =Z_{top}^{-\frac{1}{2}}(\beta_{T}%
)\sum_{N,M}%
{\displaystyle\prod\limits_{i=p+1}^{d-1}}
\exp\left[  -\frac{\beta_{T}}{2}E_{0}(N,M)\right]  \left[  \hat{n}_{i}^{\pm
}\widehat{\tilde{n}}_{i}^{\pm}\right]  ^{\left\vert n_{i}\right\vert }\left[
\hat{m}_{i}^{\pm}\widehat{\tilde{m}}_{i}^{\pm}\right]  ^{\left\vert
m_{i}\right\vert }, \label{Omega-operator}%
\end{equation}
where for negative values of $n_{i}$ and $m_{i}$ in the multi-indices $N$ and
$M$, respectively, the operators are $\hat{n}_{i}^{-}$, etc. The operator
$\hat{\Omega}\left(  \beta_{T}\right)  $ maps the topological vacuum at zero
temperature to%
\begin{align}
|\left.  0\left(  \beta_{T}\right)  \right\rangle \rangle_{top}  &
\sim\left\{  \sum_{M}\prod_{j=p+1}^{d-1}\vartheta\left(  -\frac{i\beta
_{T}R^{2}}{4\pi(\alpha^{\prime})^{\frac{3}{2}}}C_{k}^{j}m^{k};\frac{i\beta
_{T}R^{2}}{2\pi(\alpha^{\prime})^{\frac{3}{2}}}\left[  \frac{2(d-p)+22}%
{d-p-1}\right]  \right)  \right. \nonumber\\
&  \times\left.  \exp\left\{  -\frac{\beta_{T}R^{2}}{2\pi(\alpha^{\prime
})^{\frac{3}{2}}}\left[  m_{j}^{2}+\left(  B_{jk}m^{k}\right)  ^{2}\right]
\right\}  \right\}  ^{-\frac{1}{2}}\nonumber\\
&  \times\sum\limits_{N,M}\exp\left[  -\frac{\beta_{T}}{2}E_{0}\left(
N,M\right)  \right]  \left\vert N,M\right\rangle \widetilde{\left\vert
N,M\right\rangle }, \label{topological-vacuum-T}%
\end{align}
where the zero mode energy is
\begin{equation}
E_{0}\left(  N,M\right)  =\frac{R^{2}}{2(\alpha^{\prime})^{\frac{3}{2}}%
}\left[  \left(  m\right)  ^{2}+\left(  n-Bm\right)  ^{2}\right]  .
\label{Zero-mode-energy}%
\end{equation}
Thus, the solution to the thermalization of the topological sector of the
string is the relation (\ref{topological-vacuum-T}) which has been obtained
from the basic postulate of the TFD given by the equation
(\ref{thermal-vac-string}). It represents the generalization of the TFD
thermal vacuum to the topological sector of the closed strings and defines the
thermal states for the center of mass of momenta and the winding states in
${\mathbb{T}}^{d-p-1}$, respectively. Also, we have given the explicit form of
the map $\hat{\Omega}\left(  \beta_{T}\right)  $ betweeen the topological
vacuum at zero temperature and at finite temperature in terms of the algebra
(\ref{algebra-n-m}). The operator $\hat{\Omega}\left(  \beta_{T}\right)  $
represents the generalization of the thermal Bogoliubov operator $\hat
{G}(\beta_{T})$ to the topological sector.

\subsection{Thermodynamics of closed thermal string}

The thermodynamic variables of the closed thermal string can be calculated
from the partition function derived in the previous subsection. However, that
calculation cannot be applied to the thermal D-branes since in the derivation
of (\ref{part-funct-osc-1}) and (\ref{top-partition-function-1}) only the
thermal vacuum has been used. Another possibility is to calculate the
thermodynamic variables from the entropy. From the TFD relation
(\ref{thermal-vac-1}), the entropy of string is given by the expectation value
of the entropy operator in the thermal vacuum%
\begin{equation}
\frac{1}{k_{B}}S(\beta_{T})=\left\langle \left\langle 0\left(  \beta
_{T}\right)  \left\vert \hat{K}(\beta_{T})\right\vert 0\left(  \beta
_{T}\right)  \right\rangle \right\rangle , \label{string-entropy}%
\end{equation}
where $\left\vert \left.  0\left(  \beta_{T}\right)  \right\rangle
\right\rangle $ is given by the equation (\ref{thermal-vac-final}). The
analysis from the previous subsection has shown that, in the reference frame
of the center of mass, the thermal string degrees of freedom can be split in
to an oscillator factor and a topological factor%
\begin{equation}
\hat{K}(\beta_{T})=\hat{K}_{osc}(\beta_{T})+\hat{K}_{top}(\beta_{T}).
\label{factorized entropy}%
\end{equation}
According to the TFD\ method \cite{ubook}, the entropy operator corresponding
to the string oscillators $\hat{K}_{osc}(\beta_{T})$ is related to the
Bogoliubov operator (\ref{bosonic-Bogoliubov-op}) by the relation%
\begin{equation}
e^{-i\hat{G}(\beta_{T})}=e^{-\frac{1}{2}\hat{K}_{osc}(\beta_{T})}\exp\left(
\sum_{n,\mu}\hat{\alpha}_{n}^{\mu\dagger}\widehat{\tilde{\alpha}}_{n}%
^{\mu\dagger}+\hat{\beta}_{n}^{\mu\dagger}\widehat{\tilde{\beta}}_{n}^{\mu
\dag}\right)  . \label{Bogoliubov-entropy}%
\end{equation}
The explicit form of $\hat{K}_{osc}(\beta_{T})$ can be obtained from the above
equation and it is given by the following relations%
\begin{align}
\hat{K}_{osc}(\beta_{T})  &  =\hat{K}_{osc}^{%
\mathbb{R}
}(\beta_{T})+\hat{K}_{osc}^{{\mathbb{T}}}(\beta_{T}%
),\label{string-osc-entropy-op}\\
\hat{K}_{osc}^{%
\mathbb{R}
}(\beta_{T})  &  =-\sum_{a=2}^{p}\sum_{n=0}^{\infty}\left\{  \left(  \hat
{N}_{l,n}^{a}+\hat{N}_{r,n}^{a}\right)  \ln\left[  \frac{e^{-\beta_{T}n}%
Z_{n}(\beta_{T})}{e^{-\beta_{T}n}Z_{n}(\beta_{T})+1}\right]  -\ln
[e^{-\beta_{T}n}Z_{n}(\beta_{T})+1]\right\}  ,\label{string-osc-entr-op-R}\\
\hat{K}_{osc}^{{\mathbb{T}}}(\beta_{T})  &  =-\frac{R^{2}}{2\alpha^{\prime}%
}\sum_{j=p+1}^{d-1}\sum_{n=0}^{\infty}\left\{  \left(  \hat{N}_{l,n}^{j}%
+\hat{N}_{r,n}^{j}\right)  \ln\left[  \frac{e^{-\frac{\beta_{T}R^{2}}%
{2\alpha^{\prime}}n}Z_{n}(\frac{\beta_{T}R^{2}}{2\alpha^{\prime}})}%
{e^{-\frac{\beta_{T}R^{2}}{2\alpha^{\prime}}n}Z_{n}(\frac{\beta_{T}R^{2}%
}{2\alpha^{\prime}})+1}\right]  -\ln[e^{-\frac{\beta_{T}R^{2}}{2\alpha
^{\prime}}n}Z_{n}(\frac{\beta_{T}R^{2}}{2\alpha^{\prime}})+1]\right\}  ,
\label{string-osc-entr-op-T}%
\end{align}
where $Z_{n}(\beta_{T})$ is the partition function of a single oscillator.
Note that the equation (\ref{string-osc-entropy-op}) defines the entropy of
the closed string oscillators only. For the reservoir, a similar operator can
be constructed by replacing the string operators with the corresponding tilde
operators. However, since we are interested in the entropy of the string, it
is not necessary to compute the entropy of the reservoir. By plugging
(\ref{string-osc-entr-op-R}) and (\ref{string-osc-entr-op-T}) into
(\ref{string-entropy}) and after some algebra, one can show that the entropy
of the closed string oscillators has the following form%
\begin{align}
S_{osc}(\beta_{T})  &  =(p-1)k_{B}\sum_{n=1}^{\infty}\left[  \beta
_{T}ne^{-\beta_{T}n}Z_{n}\left(  \beta_{T}\right)  -\ln Z_{n}\left(  \beta
_{T}\right)  \right] \nonumber\\
&  +(p+1-d)k_{B}\frac{R^{2}}{2\alpha^{\prime}}\sum_{n=1}^{\infty}\left[
\frac{\beta_{T}R^{2}ne^{-\frac{\beta_{T}R^{2}}{2\alpha^{\prime}}n}}%
{2\alpha^{\prime}}Z_{n}\left(  \frac{\beta_{T}R^{2}}{2\alpha^{\prime}}\right)
-\ln Z_{n}\left(  \frac{\beta_{T}R^{2}}{2\alpha^{\prime}}\right)  \right]
\label{string-osc-entropy-final}%
\end{align}
where $N_{n}$ is the eigenvalue of the number operator of the oscillator of
frequency $n$. The factor of two is from the contributions of the left- and
right-moving sectors, respectively. We note that at the critical radius
$R_{0}=\sqrt{2\alpha^{\prime}}$ the equation (\ref{string-osc-entropy-final})
reproduces the entropy of $d-2$ transversal massless scalar fields.

The contribution of the topological sector to the entropy of string
(\ref{string-entropy}) cannot be calculated in the same way as the entropy of
the oscillators because there is no TFD definition for the topological entropy
operator of the closed string in ${\mathbb{T}}^{d-p-1}$. Let us construct this
operator. According to the TFD principles expressed by the relation
(\ref{string-entropy}), the topological entropy is defined as%
\begin{equation}
\frac{1}{k_{B}}S_{top}(\beta_{T})=\left\langle \left\langle 0\left(  \beta
_{T}\right)  \left\vert \hat{K}_{top}(\beta_{T})\right\vert 0\left(  \beta
_{T}\right)  \right\rangle \right\rangle . \label{top-entropy-def}%
\end{equation}
This definition must be consistent with the derivation of the entropy from the
partition function which is given by the thermodynamical relation%
\begin{equation}
S_{top}(\beta_{T})=-\frac{\partial}{\partial T}\left[  k_{B}T\ln Z_{top}%
(\beta_{T})\right]  . \label{top-entropy-part-function}%
\end{equation}
By using (\ref{top-partition-function-1}) in the above relation, one can see
that the entropy operator in the topological sector of the closed string has
the form%

\begin{equation}
\hat{K}_{top}(\beta_{T})=\left(  2\pi R\right)  ^{p+1-d}\left[  \frac
{\beta_{T}}{2}\hat{H}_{top}\exp\left(  \frac{\beta_{T}}{2}\hat{H}%
_{top}\right)  +\left(  2\pi R\right)  ^{(p+1-d)}Z_{top}(\beta_{T})\ln
Z_{top}(\beta_{T})\exp\left(  \beta_{T}\hat{H}_{top}\right)  \right]  ,
\label{topological-entropy}%
\end{equation}
where
\begin{equation}
\hat{H}_{top}=\hat{H}_{0}+\frac{R^{2}}{2(\alpha^{\prime})^{\frac{3}{2}}}%
\hat{W}, \label{H-top-1}%
\end{equation}
and $\hat{H}_{0}$ represents the topological part of the Hamiltonian with the
eigenvalues $E_{0}\left(  N,M\right)  $ given by the relation
(\ref{Zero-mode-energy}). The vacuum expectation value of the operator
$\hat{K}_{top}(\beta_{T})$ reproduces correctly the topological entropy%
\begin{equation}
S_{top}(\beta_{T})=k_{B}\ln Z_{top}(\beta_{T})+\frac{k_{B}\beta_{T}}{2}%
Z_{top}^{-1}(\beta_{T})\sum_{N,M}E_{top}(N,M)e^{-\frac{\beta_{T}}{2}%
E_{top}(N,M)}, \label{string-topo-entropy-final}%
\end{equation}
where%
\begin{equation}
E_{top}(N,M)=\frac{R^{2}}{2(\alpha^{\prime})^{\frac{3}{2}}}\left[  \left(
m\right)  ^{2}+\left(  n-Bm\right)  ^{2}\right]  +\frac{R^{2}}{2(\alpha
^{\prime})^{\frac{3}{2}}}W, \label{E-top-eigenvalues}%
\end{equation}
are the eigenvalues of the operator $\hat{H}_{top}$.

The free energy of the thermal closed string can be computed as the vacuum
expectation value in the thermal vacuum of the operator%

\begin{equation}
\hat{F}(\beta_{T})=\hat{H}-\frac{1}{\beta_{T}}\hat{K}(\beta_{T}),
\label{free-energy-operator}%
\end{equation}
where $S=S_{osc}+S_{top}$ and $\hat{H}=\hat{H}^{%
\mathbb{R}
}+\hat{H}^{{\mathbb{T}}}$ defined by the equations (\ref{Hamiltonian-R}) and
(\ref{Hamiltonian-T}), respectively. Calculation of the vacuum expectation
value of $\hat{F}$ in the thermal vacuum gives%
\begin{equation}
F(\beta_{T})=\beta_{T}^{-1}\left[  (p-1)\sum_{n=1}^{\infty}\ln Z_{n}\left(
\beta_{T}\right)  +(d-p-1)\frac{R^{2}}{2\alpha^{\prime}}\sum_{n=1}^{\infty}\ln
Z_{n}\left(  \frac{\beta_{T}R^{2}}{2\alpha^{\prime}}\right)  +\ln
Z_{top}\left(  \beta_{T}\right)  .\right]  \label{free-energy-string}%
\end{equation}
Again, we see that at the critical radius $\sqrt{2\alpha^{\prime}}$ the free
energy is the sum of the free energy of $d-2$ massless scalar fields and the
topological free energy of the zero modes. For the maximal flat subspace
$d=p+1$, i. e. in the absence of the compact directions, the free energy is
obtained from the oscillators since in the last term all $n_{i}$ and $m_{i}$
are zero.

\section{Thermal magnetized D-branes}

In this section, we are going to derive the magnetized thermal D-brane states
in $\mathbb{R}^{p-1}\times{\mathbb{T}}^{d-p-1}$ by applying the generalized
TFD formalism derived in the previous section. The D-brane boundary states at
finite temperature in flat spacetime were defined in \cite{ivv1} as the states
from the thermal closed string Hilbert space that satisfy the same Dirichlet
and Neumann boundary conditions as the ones at zero temperature. This ammounts
to imposing two sets of boundary conditions: one for the string degrees of
freedom and the other for the reservoir degrees of freedom. The same
definition can be apply in $\mathbb{R}^{p-1}\times{\mathbb{T}}^{d-p-1}$ and
the thermal magnetized D-brane states can be calculated. From the thermal
boundary state we can calculate the entropy and the free energy of the
magnetized D-brane at finite temperature as the expectation value of the
entropy operator and the free energy operator, respectively, in the thermal
magnetized D-brane state.

\subsection{Thermal magnetized D-branes states}

The thermalization of the closed string is described by the operators
$\hat{\Omega}\left(  \beta_{T}\right)  $ and $\hat{G}(\beta_{T})$ which map
the total system from zero temperature to the finite temperature which implies
that the generic string operators $\hat{O}$ transform to $\hat{O}(\beta_{T})$
as%
\begin{equation}
\hat{O}\longrightarrow\hat{O}(\beta_{T})=\left[  \hat{\Omega}\left(  \beta
_{T}\right)  \otimes e^{i\hat{G}(\beta_{T})}\right]  \hat{O}\left[
e^{-i\hat{G}(\beta_{T})}\otimes\hat{\Omega}^{-1}\left(  \beta_{T}\right)
\right]  . \label{transform-O-G-Omega}%
\end{equation}
In the flat spacetime, the thermal D-brane boundary states have been defined
by two sets of boundary conditions obtained from the total lagrangian at
finite temperature and imposed on the Hilbert space and on the tilde Hilbert
space, respectively \cite{ivv1}. We can generalize this definition to the
magnetized D-branes in $\mathbb{R}^{p-1}\times{\mathbb{T}}^{d-p-1}$ and look
for the solutions $\left\vert \left.  B(\beta_{T})\right\rangle \right\rangle
$ of the following equations%
\begin{align}
\hat{p}^{a}(\beta_{T})\left\vert \left.  B(\beta_{T})\right\rangle
\right\rangle  &  =0,\label{bc-D-T-top-1}\\
\left(  \hat{n}_{i}-2\pi\alpha^{\prime}qF_{ij}\hat{m}^{j}\right)  (\beta
_{T})\left\vert \left.  B(\beta_{T})\right\rangle \right\rangle  &
=0,\label{bc-D-T-top-2}\\
\left[  \left(  \mathbf{1}+\mathcal{B}\right)  _{ab}\hat{\alpha}_{n}^{b}%
(\beta_{T})+\left(  \mathbf{1}+\mathcal{B}\right)  _{ab}^{T}\hat{\beta}%
_{n}^{b\dag}(\beta_{T})\right]  (\beta_{T})\left\vert \left.  B(\beta
_{T})\right\rangle \right\rangle  &  =0,\label{bc-D-T-osc-1}\\
\left[  \left(  \mathbf{1}+\mathcal{B}\right)  _{ab}\hat{\alpha}_{n}^{b\dag
}(\beta_{T})+\left(  \mathbf{1}+\mathcal{B}\right)  _{ab}^{T}\hat{\beta}%
_{n}^{b}(\beta_{T})\right]  (\beta_{T})\left\vert \left.  B(\beta
_{T})\right\rangle \right\rangle  &  =0,\label{bc-D-T-osc-2}\\
\left(  \mathcal{E}_{ij}\hat{\beta}_{n}^{j}(\beta_{T})+\mathcal{E}_{ij}%
^{T}\hat{\alpha}_{n}^{j\dag}(\beta_{T})\right)  \left\vert \left.  B(\beta
_{T})\right\rangle \right\rangle  &  =0,\label{bc-D-T-osc-3}\\
\left(  \mathcal{E}_{ij}\hat{\beta}_{n}^{j\dag}(\beta_{T})+\mathcal{E}%
_{ij}^{T}\hat{\alpha}_{n}^{j}(\beta_{T})\right)  \left\vert \left.
B(\beta_{T})\right\rangle \right\rangle  &  =0, \label{bc-D-T-osc-4}%
\end{align}
for all $n>0$. Here, the operators at finite temperature have been obtained by
applying the map (\ref{transform-O-G-Omega}) to the boundary operators from
the equations (\ref{bc-R-T-momentum}) and (\ref{bc-R-boson-1}%
)-(\ref{bc-T-boson-2}). Similar equations should be written for the tilde
operators. In order to solve the above set of equations, we firstly note that
the solution can be factorized as%
\begin{equation}
\left\vert \left.  B(\beta_{T})\right\rangle \right\rangle =\left\vert \left.
B(\beta_{T})\right\rangle \right\rangle _{top}\left\vert \left.  B(\beta
_{T})\right\rangle \right\rangle _{osc}. \label{factorized-D-T}%
\end{equation}
Then one can easily show that $\left\vert \left.  B(\beta_{T})\right\rangle
\right\rangle _{top}$ and $\left\vert \left.  B(\beta_{T})\right\rangle
\right\rangle _{osc}$ can be written in terms of zero temperature D-branes as%
\begin{align}
\left\vert \left.  B(\beta_{T})\right\rangle \right\rangle _{top}  &
=\hat{\Omega}\left(  \beta_{T}\right)  \left\vert \left.  B\right\rangle
\right\rangle _{top}^{{\mathbb{T}}}=\hat{\Omega}\left(  \beta_{T}\right)
\left\vert B\right\rangle _{top}^{{\mathbb{T}}}\widetilde{\left\vert
B\right\rangle }_{top}^{{\mathbb{T}}},\label{D-T-top}\\
\left\vert \left.  B(\beta_{T})\right\rangle \right\rangle _{osc}  &
=e^{i\hat{G}(\beta_{T})}\left\vert \left.  B\right\rangle \right\rangle
_{osc}=e^{i\hat{G}(\beta_{T})}\left\vert B\right\rangle _{osc}\widetilde
{\left\vert B\right\rangle }_{osc}, \label{D-T-osc}%
\end{align}
where $\left\vert \left.  B\right\rangle \right\rangle _{top}$ and $\left\vert
\left.  B\right\rangle B\right\rangle _{osc}=\left\vert \left.  B\right\rangle
\right\rangle _{osc}^{\mathbb{R}}\otimes\left\vert \left.  B\right\rangle
\right\rangle _{osc}^{{\mathbb{T}}}$ belong to the corresponding total Hilbert
spaces at zero temperature and represent two copies of the states given in the
equations (\ref{bosonic-B-T-top}) and (\ref{bosonic-B-R-osc}) and
(\ref{bosonic-B-T-osc}), respectively. The relations (\ref{D-T-top}) and
(\ref{D-T-osc}) represent the mapping of the magnetized D-brane states from
the total Hilbert space at zero temperature to the thermal magnetized D-brane
space from the Hilbert space of the thermal closed string. Then after some
algebra, the solution to the system (\ref{bc-D-T-top-1})-(\ref{bc-D-T-osc-4})
is found to be%
\begin{align}
&  \left\vert \left.  B(\beta_{T})\right\rangle \right\rangle =N_{p-1}%
^{2}\left(  \mathcal{B}\right)  N_{d-p-1}^{2}\left(  \mathcal{B}\right)
\prod_{i=p+1}^{d-1}\sum_{n_{i}}\delta_{n_{i}-2\pi\alpha^{\prime}qF_{i}^{k_{i}%
}m_{k_{i}}}\left[  \hat{n}_{i}^{\pm}\widehat{\tilde{n}}_{i}^{\pm}\right]
^{\left\vert n_{i}\right\vert }(\beta_{T})\left[  \hat{m}_{i}^{\pm}%
\widehat{\tilde{m}}_{i}^{\pm}\right]  ^{\left\vert m_{i}\right\vert }%
(\beta_{T})\nonumber\\
&  \otimes\left(  {\displaystyle\prod\limits_{n=1}^{\infty}}\exp\left\{
-\left[  \hat{\alpha}_{n}^{a\dag}(\beta_{T})M_{ab}\hat{\beta}_{n}^{b\dag
}(\beta_{T})+\widehat{\tilde{\alpha}}_{n}^{a\dag}(\beta_{T})\tilde{M}%
_{ab}\widehat{\tilde{\beta}}_{n}^{b\dag}(\beta_{T})\right]  \right\}  \right)
\left\vert \left.  0(\beta_{T})\right\rangle \right\rangle _{osc}^{\mathbb{R}%
}\nonumber\\
&  \otimes\left(  {\displaystyle\prod\limits_{n=1}^{\infty}}\exp\left\{
-\frac{R^{2}}{\alpha^{\prime}}\left[  \hat{\alpha}_{n}^{i\dag}(\frac{\beta
_{T}R^{2}}{2\alpha^{\prime}})\mathcal{S}_{ij}\hat{\beta}_{n}^{j\dag}%
(\frac{\beta_{T}R^{2}}{2\alpha^{\prime}})+\widehat{\tilde{\alpha}}_{n}^{i\dag
}(\frac{\beta_{T}R^{2}}{2\alpha^{\prime}})\mathcal{S}_{ij}\widehat
{\tilde{\beta}}_{n}^{j\dag}(\frac{\beta_{T}R^{2}}{2\alpha^{\prime}})\right]
\right\}  \right)  \left\vert \left.  0(\frac{\beta_{T}R^{2}}{2\alpha^{\prime
}})\right\rangle \right\rangle _{osc}^{{\mathbb{T}}}, \label{D-T-sol-1}%
\end{align}
where $\mathcal{S}_{ij}\mathcal{=}\left(  \mathcal{E}^{-1}\right)  _{i}%
^{l}\left(  \mathcal{E}^{T}\right)  _{lj}$. The above relation represents the
thermal magnetized D-brane. It shows that the magnetized D-brane at finite
temperature is a vector from the thermal total Hilbert space. In the absence
of the topological sector and with $d=p+1$ it reduces to the already known
D-branes at finite temperature \cite{ivv1}\cite{ivv2}. Note that the
background fields from (\ref{D-T-sol-1}) are not thermalized, i. e. the
average values of the corresponding string states are taken to be the same as
at zero temperature. Therefore, one can take $M_{ab}=\tilde{M}_{ab}$ and
$\mathcal{E}_{ij}=\mathcal{\tilde{E}}_{ij}$. As the rest of thermal string
states, there is no simple interpretation of $\left\vert \left.  B(\beta
_{T})\right\rangle \right\rangle $ in terms of magnetized D-branes at zero temperature.

\subsection{Thermodynamics of magnetized D-branes}

The thermodynamic properties of the magnetized D-branes at finite temperature
can be derived from their entropy. The entropy and the free energy of the
thermal D-branes are defined as the expectation values of $\hat{K}$ and
$\hat{F}$ operators from the equations (\ref{factorized entropy}) and
(\ref{free-energy-operator}), respectively, in the state $\left\vert \left.
B(\beta_{T})\right\rangle \right\rangle $%
\begin{equation}
\frac{1}{k_{B}}S_{D}(\beta_{T})=\left\langle \left\langle B\left(  \beta
_{T}\right)  \left\vert \hat{K}(\beta_{T})\right\vert B\left(  \beta
_{T}\right)  \right\rangle \right\rangle . \label{Def-Entropy-D-osc}%
\end{equation}
Then one can write%
\begin{align}
S_{D}(\beta_{T})  &  =S_{D}(\beta_{T})_{osc}+S_{D}(\beta_{T})_{top}%
,\label{sum-Entropy-D-osc-1}\\
S_{D}(\beta_{T})_{osc}  &  =S_{D}(\beta_{T})_{osc}^{\mathbb{R}}+S_{D}%
(\beta_{T})_{osc}^{{\mathbb{T}}}, \label{sum-Entropy-D-osc-2}%
\end{align}
The computations are somewhat lengthy but straightforward. The main technical
detail is that the Bogoliubov transformations can be linearized \cite{ubook}%
\begin{align}
\hat{\alpha}_{n}^{a}  &  =\sqrt{Z_{n}\left(  \beta_{T}\right)  }\hat{\alpha
}_{n}^{a}\left(  \beta_{T}\right)  +\sqrt{Z_{n}\left(  \beta_{T}\right)
-1}\widehat{\tilde{\alpha}}_{n}^{a\dagger}\left(  \beta_{T}\right)  ,\\
\hat{\alpha}_{n}^{a\dagger}  &  =\sqrt{Z_{n}\left(  \beta_{T}\right)  }%
\hat{\alpha}_{n}^{a\dagger}\left(  \beta_{T}\right)  +\sqrt{Z_{n}\left(
\beta_{T}\right)  -1}\widehat{\tilde{\alpha}}_{n}^{a}\left(  \beta_{T}\right)
, \label{Bogoliubov-linear}%
\end{align}
with similar relations holding for the $\hat{\beta}$ operators and tilde
operators. Note that in ${\mathbb{T}}^{d-p-1}$, $\beta_{T}$ should be replaced
by $\beta_{T}R^{2}/2\alpha^{\prime}$. The contribution of the oscillators from
$\mathbb{R}^{p-1}$ to the entropy of the magnetized D-brane is%
\begin{align}
&  S_{D}(\beta_{T})_{osc}^{\mathbb{R}}=-2V_{D,top}^{{\mathbb{T}}}%
V_{D,osc}^{{\mathbb{T}}}k_{B}\times\nonumber\\
&  \left\{  \sum_{n=1}^{\infty}\sum_{a=2}^{p}\left[
{\displaystyle\prod\limits_{m\neq n=1}^{\infty}}
{\displaystyle\prod\limits_{c\neq a=2}^{p}}
{\displaystyle\prod\limits_{d=2}^{p}}
\left[  \sum_{k_{cd}^{m}=0}^{\infty}\left(  M_{cd}\right)  ^{2k_{cd}^{m}%
}\right]  \right]  \right. \nonumber\\
&  \times\left[  2Z_{n}\left(  \beta_{T}\right)  -1\right]  \ln\left(
\frac{e^{-\beta_{T}n}Z_{n}\left(  \beta_{T}\right)  }{e^{-\beta_{T}n}%
Z_{n}\left(  \beta_{T}\right)  +1}\right)  \left[  \sum_{s_{ad}^{n}=0}%
^{\infty}s_{ad}^{n}\left(  M_{cd}\right)  ^{2s_{ad}^{n}}\right] \nonumber\\
&  \left.  +(p-1)V_{D,osc}^{\mathbb{R}}\sum_{n=1}^{\infty}\left\{  \left[
Z_{n}\left(  \beta_{T}\right)  -1\right]  \ln\left(  \frac{e^{-\beta_{T}%
n}Z_{n}\left(  \beta_{T}\right)  }{e^{-\beta_{T}n}Z_{n}\left(  \beta
_{T}\right)  +1}\right)  +\ln\left[  e^{-\beta_{T}n}Z_{n}\left(  \beta
_{T}\right)  +1\right]  \right\}  \right\}  . \label{Entropy-D-osc-R}%
\end{align}
Here, we have denoted by $V_{D,top}^{{\mathbb{T}}}$, $V_{D,osc}^{{\mathbb{T}}%
}$ and $V_{D,osc}^{\mathbb{R}}$ the norms of $\left\vert \left.  B\left(
\beta_{T}\right)  \right\rangle \right\rangle _{top}^{{\mathbb{T}}}$,
$\left\vert \left.  B\left(  \beta_{T}\right)  \right\rangle \right\rangle
_{osc}^{{\mathbb{T}}}$ and $\left\vert \left.  B\left(  \beta_{T}\right)
\right\rangle \right\rangle _{osc}^{\mathbb{R}}$, respectively. Since the
Bogoliubov operator is unitary, the norm of the oscillator boundary states is
the square of the of the norm of the corresponding boundary states at zero
temperature. The terms $\left(  M_{cd}\right)  ^{2k_{cd}^{m}}$ represent the
matrix elements of $M_{ab}=\tilde{M}_{ab}$ at the corresponding power. The
overall factor of 2 in front of all terms is a result of adding the entropy of
the left-moving and the right-moving oscillators. The entropy of the
oscillators in ${\mathbb{T}}^{d-p-1}$ can be calculated in the same way as
$S_{D}(\beta_{T})_{osc}^{\mathbb{R}}$ and the result is formally the same
\begin{align}
&  S_{D}(\beta_{T})_{osc}^{{\mathbb{T}}}=-2V_{D,top}^{\mathbb{R}}%
V_{D,osc}^{\mathbb{R}}k_{B}\times\nonumber\\
&  \left\{  \sum_{n=1}^{\infty}\sum_{i=p+1}^{d-1}\left[
{\displaystyle\prod\limits_{m\neq n=1}^{\infty}}
{\displaystyle\prod\limits_{l\neq i=p+1}^{d-1}}
{\displaystyle\prod\limits_{r=p+1}^{d-1}}
\left[  \sum_{k_{lr}^{m}=0}^{\infty}\left(  \mathcal{S}_{lr}\right)
^{2k_{lr}^{m}}\right]  \right]  \right. \nonumber\\
&  \times\left[  2Z_{n}\left(  \frac{\beta_{T}R^{2}}{2\alpha^{\prime}}\right)
-1\right]  \ln\left(  \frac{e^{-\frac{\beta_{T}R^{2}}{2\alpha^{\prime}}n}%
Z_{n}\left(  \frac{\beta_{T}R^{2}}{2\alpha^{\prime}}\right)  }{e^{-\frac
{\beta_{T}R^{2}}{2\alpha^{\prime}}n}Z_{n}\left(  \frac{\beta_{T}R^{2}}%
{2\alpha^{\prime}}\right)  +1}\right)  \left[  \sum_{t_{ir}^{n}=0}^{\infty
}t_{ir}^{n}\left(  \mathcal{S}_{lr}\right)  ^{2t_{lr}^{m}}\right] \nonumber\\
&  +(d-p-1)V_{D,osc}^{{\mathbb{T}}}\sum_{n=1}^{\infty}\left[  Z_{n}\left(
\frac{\beta_{T}R^{2}}{2\alpha^{\prime}}\right)  -1\right]  \ln\left(
\frac{e^{-\frac{\beta_{T}R^{2}}{2\alpha^{\prime}}n}Z_{n}\left(  \frac
{\beta_{T}R^{2}}{2\alpha^{\prime}}\right)  }{e^{-\frac{\beta_{T}R^{2}}%
{2\alpha^{\prime}}n}Z_{n}\left(  \frac{\beta_{T}R^{2}}{2\alpha^{\prime}%
}\right)  +1}\right) \nonumber\\
&  \left.  +(d-p-1)V_{D,osc}^{{\mathbb{T}}}\sum_{n=1}^{\infty}\ln\left[
e^{-\frac{\beta_{T}R^{2}}{2\alpha^{\prime}}n}Z_{n}\left(  \frac{\beta_{T}%
R^{2}}{2\alpha^{\prime}}\right)  +1\right]  \right\}  .
\label{Entropy-D-osc-T}%
\end{align}

The contribution of the topological sector to the entropy can be calculated
from the relations (\ref{topological-entropy}) and (\ref{D-T-sol-1}) and the
result is%
\begin{align}
&  S_{D}(\beta_{T})_{top}=R^{p+1-d}\alpha^{\prime\frac{d-p-1}{2}}%
V_{D,osc}^{\mathbb{R}}V_{D,osc}^{{\mathbb{T}}}k_{B}Z_{top}^{-1}\left(
\beta_{T}\right)  \times\nonumber\\
&  \left\{  \sum_{N,M}^{\infty}\sum_{K,S}^{p+1}\left(  2\pi R\right)
^{p+1-d}Z_{top}\left(  \beta_{T}\right)  \ln Z_{top}\left(  \beta_{T}\right)
\exp\left[  \beta_{T}E_{top}(N\pm K,M\pm S)\right]  \right. \nonumber\\
&  +\left.  \frac{\beta_{T}}{2}E_{top}(N\pm K,M\pm S)\exp\left[  \frac
{\beta_{T}}{2}E_{top}(N\pm K,M\pm S)\right]  \right\}  . \label{Entropy-D-top}%
\end{align}
The sum of the entropies (\ref{Entropy-D-osc-R}), (\ref{Entropy-D-osc-T}) and
(\ref{Entropy-D-top}) represents the entropy of the magnetized D-brane at
finite temperature in $\mathbb{R}^{p-1}\times{\mathbb{T}}^{d-p-1}$. It can be
used to derive the rest of the thermodynamic potentials of the D-brane in the
usual fashion since the thermal D-brane is at the thermodynamical equilibrium.
This is left as an exercise to the reader.

\section{Conclusions}

In this paper, we have constructed the thermal magnetized D-brane boundary
states in $\mathbb{R}^{p-1}\times{\mathbb{T}}^{d-p-1}$ and derived their
entropy. This represents a generalization of the previous results from
\cite{ivv1,ivv2} and \cite{ivv5} where the thermal bosonic and GS D-branes
were constructed and their entropy was calculated in the flat spacetime. In
order to obtained the thermal boundary states, we have generalized the
TFD\ formalism to include the zero-mode sector of the closed strings in
${\mathbb{T}}^{d-p-1}$ and we have used this generalization to obtain the
entropy and the free energy of the closed string at finite temperature. The
generalization of the TFD formalism is an interesting result by itself, since
it extends the method of the canonical thermalization to non-trivial boundary
conditions and generalizes the previous studies from \cite{skms,qskmrs} that
establish the form of the Bogoliubov operator for a scalar and spinor field on
${\mathbb{T}}^{d-p-1}$ but without the winding conditions which are specific
to the bosonic string fields. With these boundary conditions, the map from the
total Hilbert space at zero temperature to the Hilbert space at finite
temperature must be generalized to the direct product given in $\hat{\Omega
}\left(  \beta_{T}\right)  \otimes e^{i\hat{G}(\beta_{T})}$. The operator
$\hat{\Omega}\left(  \beta_{T}\right)  $ takes the states from the zero mode
sector to finite temperature and allows one to extend the TFD method to the
topological sector.

For the future directions it is an interesting subject to study the
generalization of the results from this paper to the thermalized magnetized
D-branes obtained from the supersymmetric magnetized D-branes at finite
temperature. While the construction of the thermal D-branes from
supersymmetric D-branes has been carried out in \cite{ivv5} in the flat
spacetime and with no background fields in the GS approach, the thermalization
of the supersymmetric D-branes in the RNS formalism is still an open problem
even in the flat spacetime.

The analysis of the thermodynamical properties of the thermal magnetized
D-branes in detail is a very challenging problem. Due to the presence of
infinite string modes and of the multiplication by the norm of the D-brane
state and the normalization constants $N_{p-1}\left(  \mathcal{B}\right)  $
and $N_{d-p-1}\left(  \mathcal{B}\right)  $, several quantities are divergent
and should be renormalized before extracting any physical information from
them. This is a general problem of the boundary state description of the
D-branes at zero or finite temperature. Nevertheless, interesting information
could be obtained at a finite energy scale where only a finite number of
string modes contribute. This truncation could be helpful to the analysis of
the thermodynamics of the magnetized D-branes as a function of radius which is
an interesting problem, too, and which requires a carefully treatment of the
T-duality of the compact directions at finite temperature.

\section*{Acknowledgments}

The authors would like to acknowledge to the anonymous referee for the
positive feedback and for pointing out an important error and several
misprints in the first version of this paper. M. A. S. and I. V. V. would like
to thank to J. A. Helay\"{e}l-Neto and A. M. O. de Almeida for hospitality at
LAFEX-CBPF where part of this work was accomplished. R. N. would like to thank
to J. A. Helay\"{e}l-Neto for useful discussions and acknowledges a FAPERJ
fellowship. The research of I. V. V. has been partially supported by FAPERJ
Grant E-26/110.099/2008.


\end{document}